\documentclass[12pt]{article}

\usepackage{latexsym}
\usepackage{amssymb,amsfonts,amsmath}
\usepackage{graphicx} 
\usepackage{indentfirst}
\usepackage{bbm}
\usepackage{amssymb}
\usepackage{verbatim}
\usepackage{amsmath, amsthm,amssymb}
\usepackage{mathrsfs}
\usepackage{hyperref}
\usepackage{amsfonts}
\usepackage{dsfont}

\parskip=\medskipamount

\newcommand {\cD}{{\cal D}}
\newcommand {\cE}{{\cal E}}

\newcommand {\cN}{{\cal N}}

\newcommand {\cS}{{\cal S}}

\newcommand {\cV}{{\cal V}}

\def\a{\alpha}

\def\b{\beta}

\def\d{\delta}

\def\g{\gamma}
\def\G{\Gamma}

\def\k{\kappa}
\def\l{\lambda}
\def\m{\mu}

\def\q{\theta}
\def\r{\rho}

\def\t{\tau}

\def\x{\xi}
\def\z{\zeta}

\def\F{\Phi}
\def\J{\Psi}

\def\O{\Omega}

\def\rd{{\rm d}}
\def\ri{{\rm i}}

\newcommand{\ve}{\varepsilon}                            
\newcommand{\cDB}{{\bar\cD}}                            

\newcommand{\pa}{\partial}                           
\newcommand{\hf}{\frac12}

\newcommand{\be}{\begin{equation}}
\newcommand{\ee}{\end{equation}}
\newcommand{\bea}{\begin{eqnarray}}
\newcommand{\eea}{\end{eqnarray}}
\newcommand{\non}{\nonumber}
\def\double #1{#1{\hbox{\kern-2pt $#1$}}}

\newif\ifdtup

\def\de{{\nabla}}                                         

\newcommand{\bsubeq}{\begin{subequations}}
\newcommand{\esubeq}{\end{subequations}}

\numberwithin{equation}{section}

\begin{document}

\begin{titlepage}
\begin{flushright}
September, 2018 \\
\end{flushright}
\vspace{5mm}

\begin{center}
{\Large \bf Higher spin supermultiplets in three dimensions:
(2,0) AdS supersymmetry }
\\ 
\end{center}

\begin{center}

{\bf
Jessica Hutomo and Sergei M. Kuzenko} \\
\vspace{5mm}

\footnotesize{
{\it Department of Physics M013, The University of Western Australia\\
35 Stirling Highway, Crawley W.A. 6009, Australia}}  
~\\

\vspace{2mm}
~\\
\texttt{jessica.hutomo@research.uwa.edu.au, sergei.kuzenko@uwa.edu.au}\\
\vspace{2mm}

\end{center}

\begin{abstract}
\baselineskip=14pt

Within the framework of (2,0) anti-de Sitter (AdS) supersymmetry
in three dimensions, 
we propose a multiplet of higher-spin currents. Making use of this supercurrent, 
we construct two off-shell gauge formulations for 
a massless multiplet of half-integer superspin $(s+\frac 12)$, 
for every integer $s>0$. In the $s=1$ case, one formulation describes 
the linearised action for (2,0) anti-de Sitter supergravity, while  
the other gives the type III minimal supergravity action in (2,0) AdS superspace,
with both linearised supergravity actions originally derived  in arXiv:1109.0496.
We formulate topologically massive 
higher-spin supermultiplets in (2,0) AdS superspace. 
Our results admit a natural extension to the case of $S^3$. 

\end{abstract}
\vspace{5mm}

\vfill

\vfill
\end{titlepage}

\newpage
\renewcommand{\thefootnote}{\arabic{footnote}}
\setcounter{footnote}{0}

\tableofcontents{}
\vspace{1cm}
\bigskip\hrule

\allowdisplaybreaks

\section{Introduction}

In four dimensions (4D), 
there is an interesting correspondence between 
$\cN=1$ anti-de Sitter (AdS) supergravity\footnote{Townsend's work 
on $\cN=1$ AdS supergravity  \cite{Townsend}
appeared shortly after Freedman and Das constructed 
$\cN=2$ AdS supergravity \cite{FD}. The motivations for  \cite{Townsend}
and \cite{FD} were rather different.
}
 \cite{Townsend} and massless
higher-spin supermultiplets in AdS${}_4$ \cite{KS94}.
Specifically,  two off-shell formulations are known
for pure $\cN=1$ AdS supergravity, the minimal 
\cite{FGvN,FvN2,Siegel}
(see, e.g., \cite{GGRS,BK} 
for pedagogical reviews) and the non-minimal \cite{BK12} theories. 
In AdS${}_4$ there exist two series of  massless off-shell gauge supermultiplets 
of half-integer superspin $s+\hf$, with $s=1,2,\dots$
 \cite{KS94}.\footnote{Such a supermultiplet
describes two ordinary massless spin-$(s+\hf)$ and spin-$(s+1)$ fields
  on-shell.}
The correspondence consists of the fact that,
 for the lowest superspin value corresponding to $s=1$, 
 one series yields the linearised action for
minimal  AdS supergravity, while the other leads to   
  linearised non-minimal  AdS supergravity.
It has recently been pointed out \cite{HKO} that  a similar correspondence might occur
in the case of 3D $\cN=2$ supersymmetry,  
which is a natural cousin of the 4D $\cN=1$ one.

Unlike four dimensions, where pure  $\cN=1$ AdS supergravity is unique on-shell, 
the  feature specific to three dimensions is 
the existence of two distinct $\cN=2$ AdS supergravity theories \cite{AT}, 
which are known as the (1,1) and (2,0) AdS supergravity theories, 
originally constructed as Chern-Simons theories. 
Two off-shell formulations for  (1,1) AdS supergravity have been developed, 
the minimal \cite{RvanN86,ZupnikPak,NG,BCSS,KLT-M11,KT-M11,KLRST-M}  and the non-minimal  \cite{KT-M11,KLRST-M} theories,  and one for (2,0) AdS supergravity
 \cite{HIPT,KLT-M11,KT-M11,KLRST-M}. 
Since there are three off-shell $\cN=2$ AdS supergravity theories, 
one might expect the existence of three  series of massless 
higher-spin gauge supermultiplets.  
In a recent paper \cite{HKO}, we have  presented two series of massless higher-spin actions which are associated with the minimal and the non-minimal (1,1) AdS 
supergravity theories, respectively, 
generalising similar constructions in the super-Poincar\'e case \cite{KO}.
The present paper is devoted to constructing higher-spin gauge multiplets with 
(2,0) AdS supersymmetry.

It is worth pointing out that the massless 3D constructions of \cite{HKO,KO}, 
were largely modelled on the 4D results of \cite{KS94,KSP}.
With respect to 3D (2,0) AdS supersymmetry, unfortunately 
there is no 4D intuition to guide us,
and new ideas are required in order to construct higher-spin gauge supermultiplets.
In this paper our approach will be to utilise an observation 
that has often been used in the past 
to formulate off-shell supergravity multiplets \cite{BdeRdeW,SohniusW1,SohniusW2,SohniusW3,HL,Howe5Dsugra}.
The idea is to make use of a higher-spin extension of the supercurrent 
(also known as the multiplet of currents), the concept introduced by Ferrara and Zumino in the case of 4D $\cN=1$ 
Poincar\'e supersymmetry \cite{FZ} and extended to 4D $\cN=2$ 
Poincar\'e supersymmetry by Sohnius \cite{Sohnius}.
Specifically, for a simple supersymmetric model in (2,0) AdS superspace 
we identify a multiplet of conserved higher-spin currents. 
In general, the multiplet of currents is always off-shell. 
Using the constructed higher-spin supercurrent, we may identify  a corresponding 
supermultiplet of higher-spin fields. The procedure to follow is concisely described
by Bergshoeff et {\it al.} \cite{BdeRdeW}:
``One first assigns a field to each component of the current multiplet,
and forms a generalized inner product of field and current components.''

Our multiplet of currents is described by the conservation equations
\begin{subequations} \label{1.1}
\bea
\cD^\b J_{\b \a_1 \dots \a_{2s-1}} = \cD_{(\a_1} {\mathbb T}_{\a_2 \dots \a_{2s-1})}~, 
\qquad \bar \cD^\b J_{\b \a_1 \dots \a_{2s-1}} 
= \bar \cD_{(\a_1} \bar {\mathbb T}_{\a_2 \dots \a_{2s-1})}~.
\eea
Here $\cD_\a $ and $\bar \cD_\a$ are the covariant spinor derivatives
of (2,0) AdS superspace \cite{KT-M11}, 
$J_{\a(2s)} := J_{\a_1 \dots \a_{2s}} = J_{(\a_1 \dots \a_{2s})} = \bar J_{\a(2s)} $
denotes  the higher-spin supercurrent, and  ${\mathbb T}_{\a(2s-2)} $  
the corresponding trace supermultiplet constrained to be 
covariantly linear\footnote{We make use of the blackboard bold letters for 
covariantly linear superfields, in accordance with the notation adopted in 
\cite{KT-M11}.}
\bea
\bar \cD^2 {\mathbb T}_{\a(2s-2)}=0~, 
\qquad \cD^2 {\mathbb T}_{\a(2s-2)}=0~.\label{1.1b}
\eea
In general, the trace supermultiplet is complex, 
\bea
{\mathbb T}_{\a(2s-2)} = {\mathbb Y}_{\a(2s-2)} -\ri {\mathbb Z}_{\a(2s-2)}~, 
\quad 
{\rm Im} \,{\mathbb Y}_{\a(2s-2)}=0~,
\quad 
{\rm Im} \,{\mathbb Z}_{\a(2s-2)}=0~.
\eea
\end{subequations} 
In the $s=1$ case, the above conservation equation coincides with that 
for the (2,0) AdS supercurrent \cite{KT-M11}.

Our work may have various generalisations and applications. 
For instance, the massless higher-spin actions 
constructed in section 4.1 are expected to  possess nonlinear completions, 
say, in the spirit of the bosonic Chern-Simons constructions of 
\cite{Blencowe,HR,CFPT}. Our results admit a natural extension to the case of $S^3$, 
which may lead to higher-spin applications of the localisation techniques, 
see, e.g., \cite{PZ,Willett} for reviews. The adequate superspace setting to 
formulate $\cN=2$ supersymmetric theories 
on $S^3$ has been developed \cite{SS}.

This paper is organised as follows. Section 2 provides a brief review of 
 (2,0) AdS superspace. In section 3 we consider simple models for 
 a chiral scalar supermultiplet and demonstrate how the higher-spin supercurrent 
 \eqref{1.1} emerges. In section 4 we develop two off-shell formulations for 
a massless multiplet of half-integer superspin $(s+\hf)$ in (2,0) AdS superspace, 
with $s$ a positive integer. Our results and their implications and possible
extensions are discussed in section 5. In the appendix we collect important 
(2,0) AdS identities. 
 

\section{(2,0) AdS superspace} \label{section2}

In this section we give a summary of the most important results concerning 
 (2,0) AdS superspace, see \cite{KT-M11} for the details.

The covariant derivatives of (2,0) AdS superspace have the form
\bea
\cD_{{A}}=(\cD_{{a}}, \cD_{{\a}},\bar \cD^\a)
=E_{{A}}+\O_{{A}}+\ri \F_{{A}} J~.
\label{CovDev}
\eea
Here $E_A$ and $\O_A$ denote
 the inverse supervielbein and 
 the Lorentz connection, respectively,
\bea
E_A=E_A{}^M \frac{\pa}{\pa z^M}~,
\qquad \O_A=\hf\O_{A}{}^{bc} M_{bc}= -\O_{A}{}^b M_b
=\hf\O_{A}{}^{\b\g}M_{\b\g}~.
\eea
The Lorentz generators with two vector indices 
($M_{ab}= -M_{ba}$), with one vector index ($M_a$)
and with two spinor indices ($M_{\a\b} = M_{\b\a} $) 
are defined in the appendix.
The ${\rm U(1)}_R$ generator $J$  in \eqref{CovDev}
is defined to act on the covariant derivatives
as follows:
\bea
{[} J,\cD_{\a}{]}
=\cD_{\a}~,
\qquad
{[} J,\cDB^{\a}{]}
=-\cDB^\a~,
\qquad 
{[}J,\cD_a{]}=0~.
\eea
The covariant derivatives satisfy the following algebra \cite{KT-M11}:
\begin{subequations} \label{deriv20}
\bea
\{\cD_\a,\cD_\b\}
&=&
0
~,\qquad
\{\cDB_\a,\cDB_\b\}
=
0~,
\\
\{\cD_\a,\cDB_\b\}
&=&
-2 \ri \cD_{\a\b} -4 \ri \ve_{\a\b} \cS J +4 \ri \cS M_{\a \b}~,\\
{[}\cD_{a},\cD_\b {]}
&=&(\g_a)_\b{}^\g \cS \cD_{\g}~, \quad
{[}\cD_{a},\cDB_\b{]}
= (\g_a)_\b{}^{\g}\cS \bar \cD_{\g}~, \\
{[}\cD_a,\cD_b]{}
&=&4  \ve_{abc}\cS^2 M^c ~.
\eea
\end{subequations}
Here the parameter $\mathcal{S}$ is related to the AdS scalar curvature as $R= -24 \mathcal{S}^2$. 

In accordance with the general formalism of \cite{BK},
the isometries of (2,0) AdS superspace are generated by  those real  supervector fields 
$\z^A E_A$ 
which obey the superspace Killing  equation \cite{KT-M11}
\begin{subequations}\label{Killing}
\bea
\big{[}\z+\ri \t J+\hf l^{bc} M_{bc},\cD_A\big{]}=0~,
\eea
where 
\bea
\z= \z^B \cD_B =\z^b\cD_b+\z^\b\cD_\b+\bar \z_\b\bar \cD^\b~,
\qquad \overline{\z^b}=\z^b~,
\eea 
\end{subequations}
and $\t$ and $l^{bc}$ are some local U(1)${}_R$ and Lorentz parameters, respectively.
Every solution of \eqref{Killing} is called a Killing supervector field of (2,0) 
AdS superspace.
As demonstrated in \cite{KT-M11}, eq. \eqref{Killing}
implies that the parameters $\z_\a$, $\t$  and $l_{\a\b}$ 
are uniquely expressed in terms of the vector  $\z_{\a\b}$, 
\bea
\z_\a= \frac{\ri}{6} \bar \cD^\b \z_{\b\a}~,\quad 
\t =\frac{\ri}{2} \cD^\a\z_\a 
~, \qquad 
l_{\a\b} =2\big( \cD_{(\a } \z_{\b)} -\cS \z_{\a\b} \big) ~,
\eea 
which obeys the equation 
\bea
\cD_{(\a}\z_{\b\g)}=0~.
\eea
It follows that $\z_a$ is a Killing vector field,
\bea
\cD_a \z_b + \cD_b \z_a=0~.
\eea
One may also prove the following relations
\bea
 \bar \cD_\a \t
=\frac{\ri}{3}\bar \cD^\b l_{\a\b} = 4\cS \z_\a~,
\qquad 
\bar \cD_\a\z_\b=0~, \qquad 
\cD_{(\a} l_{\b\g)}=0~.
\eea
The  Killing supervector fields of (2,0) AdS superspace
generate the supergroup 
$\rm OSp(2|2;{\mathbb R}) \times Sp(2,{\mathbb R}) $, the isometry group 
of (2,0) AdS superspace.
Rigid supersymmetric field theories on (2,0) AdS superspace are invariant under 
the isometry transformations. The isometry transformation associated with 
the Killing supervector field  $\z^A E_A$ acts on a tensor superfield $U$ (with its indices suppressed) by 
the rule
\bea
\d_\z U = \big(\z+\ri \t J+\hf l^{bc} M_{bc}\big)U~.
\eea
Associated with a real scalar superfield $L$ is the following 
supersymmetric  invariant
\bea
\int \rd^3x \rd^2 \q  \rd^2 \bar \q
\,E\,{ L} &=& 
-\frac{1}{4} \int
\rd^3x \rd^2 \q  
\, \cE\, {\bar \cD}^2  {L} ~, \qquad
E^{-1}= {\rm Ber}\, (E_{\rm A}{}^M)~,
\eea
where 
$\cE$ denotes the chiral integration measure.


\section{Higher-spin supercurrents for chiral matter} \label{section3}

In this section we study higher-spin supercurrents in simple models 
for a chiral scalar supermultiplet
in $(2,0)$ AdS superspace. 

\subsection{Massless models}

We first consider a massless model. Its action
\bea
S = \int \rd^3x \rd^2 \q  \rd^2 \bar \q
\,E\, \bar \F \F ~, \qquad \bar \cD_\a \F =0
\label{chiral}
\eea
is invariant under the isometry transformations of (2,0) AdS superspace
for any U(1)${}_R$ charge $w$ of the chiral superfield, 
\bea
{J} \F =- w \F ~.
\eea
The action is superconformal provided  $w=\hf$.

As in \cite{HKO}, it is useful to introduce auxiliary real variables $\z^\a \in {\mathbb R}^2$. 
Given a tensor superfield $U_{\a(m)}$, we associate with it 
the following  field
\bea
U_{(m)} (\z):= \z^{\a_1} \dots \z^{\a_m} U_{\a_1 \dots \a_m}~,
\label{e1}
\eea
which is a homogeneous polynomial of degree $m$ in the variables $\z^\a$.
We introduce operators that  increase the degree 
of homogeneity in the variable $\z^\a$, 
\begin{subequations}
\bea
{\cD}_{(1)} &:=& \z^\a \cD_\a~, \qquad \cD_{(1)}^2 =0~,\\
{\bar \cD}_{(1)} &:=&  \z^\a \bar \cD_\a~, \qquad \bar \cD_{(1)}^2 =0~,\\
{\cD}_{(2)} &:=& \ri \z^\a \z^\b \cD_{\a\b}~.
\eea
\end{subequations}
We also introduce two nilpotent operators that decrease the degree 
of homogeneity in the variable $\z^\a$, specifically
\begin{subequations}
\bea
\cD_{(-1)} &:=& \cD^\a \frac{\pa}{\pa \z^\a}~, \qquad \cD_{(-1)}^2 =0~,\\
\bar \cD_{(-1)}& :=& \bar \cD^\a \frac{\pa}{\pa  \z^\a}~, \qquad \bar \cD_{(-1)}^2 =0~,
\eea
\end{subequations}

Let us first consider the superconformal case, $w=\hf$.  The analysis
given in \cite{HKO} implies that  the theory possesses a
real supercurrent $J_{(2s)} =\bar J_{(2s)}$, for any positive integer $s$, 
which obeys the conservation equation
\bea
\cD_{(-1)} J_{(2s)} &=& 0~.
\label{3.6}
\eea
This supercurrent proves to have the same form as in the (1,1) AdS case
considered in \cite{HKO}.
Specifically, the higher-spin supercurrent\footnote{In the flat superspace limit, 
the supercurrent \eqref{3.1} reduces to the one constructed in \cite{NSU}.}
 is given by
\bea
J_{(2s)} &=& \sum_{k=0}^s (-1)^k
\left\{ \hf \binom{2s}{2k+1} 
{\cD}^k_{(2)} \bar \cD_{(1)} \bar \F \,\,
{\cD}^{s-k-1}_{(2)} \cD_{(1)} \F  
+ \binom{2s}{2k} 
{\cD}^k_{(2)} \bar \F \,\, {\cD}^{s-k}_{(2)} \F \right\}~.~~~
\label{3.1}
\eea
Making use of
the massless equations of motion, $\cD^2 \F = 0$, one may check that
this supermultiplet does obey the conservation equation \eqref{3.6}.

Now we turn to the non-superconformal case, $w\neq \hf$.
Direct calculations give
\begin{subequations} \label{3.2}
\bea
\cD_{(-1)} J_{(2s)} &=&  {\cD}_{(1)} {\mathbb T}_{(2s-2)}~,
\label{3.8a}
\eea
where we have denoted 
\bea
{\mathbb T}_{(2s-2)}&=& 2\ri (1-2w){\cS}(2s+1)(s+1) \sum_{k=0}^{s-1}\frac{1}{2s-2k+1} (-1)^{k} \binom{2s}{2k+1}
\non \\ 
&& 
\times {\cD}^k_{(2)} \bar \F \,\,{\cD}^{s-k-1}_{(2)} \F ~.
\eea
The trace multiplet ${\mathbb T}_{(2s-2)}$ is covariantly  linear,
\be
\bar {\cD}^2 {\mathbb T}_{(2s-2)} =0 ~, \qquad {\cD}^2 {\mathbb T}_{(2s-2)} =0 ~,
\label{3.8c}
\ee
as a consequence of the equations of motion and identities \eqref{A.2c}.
It is seen that 
${\mathbb T}_{(2s-2)}$ has non-zero real and imaginary parts,  
\bea
{\mathbb T}_{(2s-2)} = {\mathbb Y}_{(2s-2)} -\ri {\mathbb Z}_{(2s-2)}~,
\qquad \bar {\mathbb Y}_{(2s-2)} ={\mathbb Y}_{(2s-2)}~, \qquad
\bar {\mathbb Z}_{(2s-2)}={\mathbb Z}_{(2s-2)}~,
\eea
\end{subequations}
except for the  $s=1$  case which is characterised by ${\mathbb Y}=0$.
For $s=1$ the above results agree with \cite{KT-M11}.
The technical details of the derivation of \eqref{3.2} are collected in the appendix.

The above results can be used to derive higher-spin supercurrents
in a non-minimal scalar supermultiplet model described by the action
\bea
S = -\int \rd^3x \rd^2 \q  \rd^2 \bar \q
\,E\, \bar \G \G ~, \qquad \bar \cD^2  \G =0~,
\label{non-minimal}
\eea
with $\G$ being a complex linear superfield.\footnote{Unlike eq. \eqref{1.1b},
the above condition on $\G$ is the only constraint obeyed by $\G$.}
The non-minimal theory \eqref{non-minimal} proves to be dual to \eqref{chiral}
provided the U(1)${}_R$ weight of $\G$ is opposite to that of $\F$, 
\bea
{J} \G =w \G ~.
\eea
Replacing $\F \to \bar \G$ and $\bar \F \to \G$ in \eqref{3.2} gives the higher-spin 
supercurrents in the non-minimal theory \eqref{non-minimal}, 
which is similar to the 4D case \cite{KKvU,BHK}.


\subsection{Massive model}

Let us add a mass term to the functional  \eqref{chiral} and consider the following action
\bea
S = \int \rd^3x \rd^2 \q  \rd^2 \bar \q \,E\, \bar \F \F
+\Big\{ \hf \int \rd^3x \rd^2 \q \, \cE \, m\F^2 +{\rm c.c.} \Big\}~,
\label{chiral-massive}
\eea
with $m$ a complex mass parameter. 
In the $m\neq 0$ case,   the U(1)${}_R$ weight of $\F$ 
is uniquely fixed to be $w=1$, in order for the action to be $R$-invariant.

Making use of the massive equations of motion
\bea
-\frac{1}{4} \cD^2 \F  +\bar m \bar \F =0, \qquad
-\frac{1}{4} \bar \cD^2 \bar \F +m \F =0,
\eea
we obtain 
\bea
\cD_{(-1)}J_{(2s)} &=& -2\ri{\cS}(2s+1)(s+1) \cD_{(1)}
\sum_{k=0}^{s-1}\frac{1}{2s-2k+1} (-1)^{k} \binom{2s}{2k+1}
\non \\ 
&& 
\times {\cD}^k_{(2)} \bar \F \,\,{\cD}^{s-k-1}_{(2)}  \F 
\non \\
&& + \bar m \,(-1)^s (2s+1)\sum_{k=0}^{s-1}\left\{1+ (-1)^s \frac{2k+1}{2s-2k+1}\right\} (-1)^{k} \binom{2s}{2k+1}
\non \\ 
&& 
\times {\cD}^k_{(2)} \bar \F \,\,{\cD}^{s-k-1}_{(2)} \bar \cD_{(1)} \bar \F  ~,
\label{5.1}
\eea
where $J_{(2s)}$ is defined by \eqref{3.1}.
We observe that \eqref{5.1} can also be written in the form 
\bea
\cD_{(-1)}J_{(2s)} &=& \hf (-1)^s \,\cD_{(-1)} \sum_{k=0}^{s-1} (-1)^{k} \binom{2s}{2k+1} \cD^{k}_{(2)} \cD_{(1)} \F \,\, \cD^{s-k-1}_{(2)} \bar \cD_{(1)} \bar \F 
\non \\ 
&&-\hf \cD_{(1)} \sum_{k=0}^{s-1} (2k+1)(-1)^{k} \binom{2s}{2k+1}  \cD^{k}_{(2)} \cD^{\a} \F \,\, \cD^{s-k-1}_{(2)} \bar \cD_{\a} \bar \F 
\non \\
&&+ 2\ri \cS \, \cD_{(1)} \sum_{k=0}^{s-1} \left[(2k+1) + (-1)^{s-1} s(2s-2k-1) \right] 
\non \\
&&\qquad \times (-1)^{k} \binom{2s}{2k+1} \cD^{k}_{(2)}  \F \,\, \cD^{s-k-1}_{(2)}  \bar \F 
\non \\
&&+ \ri [1+ (-1)^s ] \sum_{k=0}^{s-1} (2k+1) (-1)^{k} \binom{2s}{2k+1} 
\non \\
&& \qquad \times \cD^{k}_{(2)} \cD^{\a} \F \,\, \cD^{s-k-1}_{(2)} \z^{\b} \cD_{\a \b} \bar \F ~.
\label{5.2}
\eea
Thus, for all odd values of $s$, 
\begin{subequations}
\bea
s= 2n+1~, \qquad n =0, 1, \dots~, \label{3.13a}
\eea
we end up with the conservation equation 
\bea
\cD_{(-1)} \hat{J}_{(2s)} = \cD_{(1)} \hat{\mathbb T}_{(2s-2)}
\label{3.13b}
\eea
where we have denoted 
\bea
\hat{J}_{(2s)} &=& J_{(2s)}
- \hf \sum_{k=0}^s (-1)^k
 \binom{2s}{2k+1} 
{\cD}^k_{(2)} \bar \cD_{(1)} \bar \F \,\,
{\cD}^{s-k-1}_{(2)} \cD_{(1)} \F  
~,\\
\hat{\mathbb T}_{(2s-2)} &=& -\hf \sum_{k=0}^{s-1} (2k+1)(-1)^{k} \binom{2s}{2k+1} 
 \cD^{k}_{(2)} \cD^{\a} \F \,\, \cD^{s-k-1}_{(2)} \bar \cD_{\a} \bar \F 
\non \\
&&+ 2\ri \cS \,  \sum_{k=0}^{s-1} \left[(1-s)(2k+1)+ 2s^2 \right] (-1)^{k} \binom{2s}{2k+1} 
\cD^{k}_{(2)}  \F \,\, \cD^{s-k-1}_{(2)}  \bar \F
~.~~~
\eea
The trace multiplet $\hat{\mathbb T}_{(2s-2)}$ is covariantly linear, 
\bea
\bar \cD^2 \hat{\mathbb T}_{(2s-2)} = 0 ~,\qquad \cD^2 \hat{\mathbb T}_{(2s-2)} =0~.
\label{3.13e}
\eea
\end{subequations}
The conservation equation defined by eqs. \eqref{3.13b} and \eqref{3.13e} 
coincides with that defined by eqs. \eqref{3.8a} and \eqref{3.8c}.

The above consideration demonstrates that in the massive case 
higher-spin supercurrents $ \hat{J}_{(2s)} $ exist only for the odd values of $s$, eq. \eqref{3.13a}. This conclusion  is analogous to the earlier results 
in four dimensions \cite{BGK1,HK1,BHK}. As was demonstrated \cite{BHK} in  the 4D case, 
the even values of $s$ are also allowed provided there are several massive chiral superfields
in the theory. The analysis of \cite{BHK} may be extended to the 3D (2,0) AdS case. 


\section{Massless higher-spin gauge theories} \label{section4}

The explicit  structure of the higher-spin supercurrent defined by eqs. \eqref{3.8a}
and \eqref{3.8c} 
allows us to develop two off-shell formulations for 
a massless multiplet of half-integer superspin $(s+\hf)$, 
for every integer $s>0$. We will call them type II and type III  models 
in order to comply with the terminology introduced in \cite{KT-M11}
for the minimal formulations of $\cN=2$ supergravity.


\subsection{Type II series}

Given a positive integer $s \geq 2$, 
we propose to describe a massless multiplet of half-integer superspin $(s+\hf)$ 
in terms of the following dynamical variables:
\be
\cV^{(\rm II)}_{(s+\hf )} = 
\Big\{ {\mathfrak H}_{\a(2s)}, \mathfrak{L}_{\a(2s-2)} \Big\} ~.
\label{2.1}
\ee
Here ${\mathfrak H}_{\a(2s)} = {\mathfrak H}_{(\a_1 \dots \a_{2s})}$ 
and ${\mathfrak L}_{\a(2s-2)} = {\mathfrak L}_{(\a_1 \dots \a_{2s-2})}$
are unconstrained real tensor superfields. 
We postulate gauge transformations for the dynamical superfields:
\begin{subequations} \label{lambda-gauge}
\bea
\d_\l {\mathfrak H}_{\a(2s)}&=& 
{\bar \cD}_{(\a_1} \l_{\a_2 \dots \a_{2s})}-{\cD}_{(\a_1}\bar \l_{\a_2 \dots \a_{2s})} ~, 
\label{H-gauge} \\ 
\d_\l {\mathfrak L}_{\a(2s-2)} &=& -\frac{\ri}{2}
\big( \bar \cD^{\b} \l_{\b \a(2s-2)}+ \cD^{\b} \bar \l_{\b \a(2s-2)} \big)~,
\label{L-gauge}
\eea
\end{subequations}
where the gauge parameter $\l_{\a(2s-1)}$ is unconstrained complex.
In order for $\d_\l {\mathfrak H}_{\a(2s)}$ and $\d_\l {\mathfrak L}_{\a(2s-2)}$
to be real,  $\l_{\a(2s-1)}$ must be
charged under the $R$-symmetry group U(1)${}_{R}$:
\bea
J \l_{\a(2s-1)} = \l_{\a(2s-1)}~, \qquad J \bar \l_{\a(2s-1)}= -\bar \l_{\a(2s-1)}~.
\eea
Equation \eqref{H-gauge} is the gauge transformation law of a conformal 
superspin-$(s+\hf)$ gauge multiplet \cite{HKO}. It is natural to interpret 
${\mathfrak L}_{\a(2s-2)} $ as a compensating multiplet.

We postulate the compensator ${\mathfrak L}_{\a(2s-2)} $ to have its own 
gauge freedom of the form 
\bea
\d_\x {\mathfrak L}_{\a(2s-2)} 
=  { \x}_{\a(2s-2)}+ \bar { \x}_{\a(2s-2)} ~, \qquad  \bar \cD_{\b} \x_{\a(2s-2)}=0~,
\label{prep-gauge}
\eea
with the gauge parameter ${\x_{\a(2s-2)}}$ being covariantly chiral, 
but otherwise arbitrary. 
It should be pointed out that in (1,1) AdS superspace covariantly chiral superfields
exist only in the scalar case, since the constraint $\bar \cD_\b\J_{\a(n)}=0$ is inconsistent 
for $n>0$. Therefore, the gauge transformation law \eqref{prep-gauge}
is specific for the (2,0) AdS supersymmetry.

Associated with ${\mathfrak L}_{\a(2s-2)} $
is the real field strength 
\bea
 \mathbb{L}_{\a(2s-2)} = \ri \cD^{\b} \bar \cD_{\b} {\mathfrak L}_{\a(2s-2)} ~,
 \qquad \mathbb{L}_{\a(2s-2)}= \bar{\mathbb{L}}_{\a(2s-2)}~,
\label{2.2}
\eea
which is invariant under the gauge transformations \eqref{prep-gauge},
$\d_\x  \mathbb{L}_{\a(2s-2)} =0$. It is not difficult to see that 
$\mathbb{L}_{\a(2s-2)}$ is a covariantly linear superfield, 
\bea
{\cD}^2 \mathbb{L}_{\a(2s-2)}=0 ~.
\eea
From \eqref{L-gauge}
we can read off the gauge transformation of the field strength 
\bea
\d_\l \mathbb{L}_{\a(2s-2)}&=&
\frac{1}{4}
\big( \cD^{\b} {\bar \cD}^2 \l_{\b \a(2s-2)}- \bar \cD^{\b} {\cD}^2 \bar \l_{\b \a(2s-2)}\big)
~.~~~
\label{bbL-gauge}
\eea

Modulo an overall  normalisation factor, there is a unique quadratic action 
which is invariant under 
the gauge transformations \eqref{lambda-gauge}. It is given by 
\bea
S^{(\rm II)}_{(s+\hf)}[{\mathfrak H}_{\a(2s)} ,{\mathfrak L}_{\a(2s-2)} ]
&=& \Big(-\hf \Big)^{s} 
\int 
\rd^3x \rd^2 \q  \rd^2 \bar \q
\, E \bigg\{\frac{1}{8}{\mathfrak H}^{\a(2s)}
\cD^{\b}\bar{\cD}^{2} \cD_{\b}
{\mathfrak H}_{\a(2s)} \non \\
&&-\frac{s}{8}([\cD_{\b},\bar{\cD}_{\g}]{\mathfrak H}^{\b \g \a(2s-2)})
[\cD^{\d},\bar{\cD}^{\r}]{\mathfrak H}_{\d \r \a(2s-2)}
 \non \\
&& +\frac{s}{2}(\cD_{\b \g}{\mathfrak H}^{\b \g \a(2s-2)})
\cD^{\d \r}{\mathfrak H}_{\d \r \a(2s-2)}+ 2\ri s \,{\cS} {\mathfrak H}^{\a(2s)} {\cD}^\b {\bar \cD}_{\b} {\mathfrak H}_{\a(2s)}
\non \\
&&-  \frac{2s-1}{2} \Big( \mathbb{L}^{\a(2s-2)} [\cD^{\b}, \bar \cD^{\g}] {\mathfrak H}_{\b \g \a(2s-2)}
+ 2
\mathbb{L}^{\a(2s-2)} \mathbb{L}_{\a(2s-2)} \Big) \non\\
&& 
-\frac{(s-1)(2s-1)}{4s} \Big(  \cD_{\b} \mathfrak{L}^{\b \a(2s-3)} \bar \cD^2 \cD^{\g} \mathfrak{L}_{\g \a(2s-3)} +{\rm c.c.} \Big)
\non \\
&&-4(2s-1) \cS \mathfrak{L}^{\a(2s-2)} \mathbb{L}_{\a(2s-2)}
\bigg\}~.
\label{action}
\eea
By construction, the action is also invariant under \eqref{prep-gauge}.

Setting $s=1$ in \eqref{action} gives the linearised 
action for (2,0) AdS supergravity, which was originally derived 
in section 10.1 of \cite{KT-M11}.\footnote{Ref. \cite{KT-M11} made use of the curvature
parameter $\r$, which is related to our $\cS$ as $\r =4\cS$.}
It should be remarked that the second last term in \eqref{action} is not defined
in the  $s=1$ case. However, this term contains an overall  numerical factor $(s-1)$
and therefore it does not contribute for $s=1$.


\subsection{Type III series}

Our second model for the massless superspin-$(s+\hf)$ multiplet 
is realised in terms of dynamical variables 
that are completely similar to \eqref{2.1}, 
\be
\cV^{(\rm III)}_{(s+\hf )} = 
\Big\{ {\mathfrak H}_{\a(2s)}, \mathfrak{V}_{\a(2s-2)} \Big\} ~.
\ee
Here ${\mathfrak H}_{\a(2s)}$
and ${\mathfrak V}_{\a(2s-2)}$
are unconstrained real tensor superfields. 
The only difference from the type II case consists in a different gauge transformation
law for the compensator ${\mathfrak V}_{\a(2s-2)} $.
We postulate the following gauge transformation laws:
\begin{subequations}\label{4.10}
\bea
\d_\l {\mathfrak H}_{\a(2s)}&=& 
{\bar \cD}_{(\a_1} \l_{\a_2 \dots \a_{2s})}-{\cD}_{(\a_1}\bar \l_{\a_2 \dots \a_{2s})} ~,  \\ 
\d_\l {\mathfrak V}_{\a(2s-2)} &=& \frac{1}{2s}
\big( \bar \cD^{\b} \l_{\b \a(2s-2)}- \cD^{\b} \bar \l_{\b \a(2s-2)} \big)~,
\label{V-gauge}
\eea
\end{subequations}
where the gauge parameter $\l_{\a(2s-1)}$ is unconstrained complex.
The compensator ${\mathfrak V}_{\a(2s-2)} $ is required to have its own 
gauge freedom of the form 
\bea
\d_\x {\mathfrak V}_{\a(2s-2)} 
=  { \x}_{\a(2s-2)}+ \bar { \x}_{\a(2s-2)} ~, \qquad  \bar \cD_{\b} \x_{\a(2s-2)}=0~,
\label{prep-gauge2}
\eea
with the gauge parameter ${\x_{\a(2s-2)}}$ being covariantly chiral, 
but otherwise arbitrary. 

A unique gauge-invariant action is given by 
\bea
S^{(\rm III)}_{(s+\hf)}
&=& \Big(-\hf \Big)^{s} 
\int 
\rd^3x \rd^2 \q  \rd^2 \bar \q
\, E \bigg\{\frac{1}{8}{\mathfrak H}^{\a(2s)}
\cD^{\b}\bar{\cD}^{2} \cD_{\b}
{\mathfrak H}_{\a(2s)} \non \\
&&-\frac{1}{16}([\cD_{\b},\bar{\cD}_{\g}]{\mathfrak H}^{\b \g \a(2s-2)})
[\cD^{\d},\bar{\cD}^{\r}]{\mathfrak H}_{\d \r \a(2s-2)}
 \non \\
&& +\frac{1}{4}(\cD_{\b \g}{\mathfrak H}^{\b \g \a(2s-2)})
\cD^{\d \r}{\mathfrak H}_{\d \r \a(2s-2)}+ \ri  \,{\cS} {\mathfrak H}^{\a(2s)} {\cD}^\b {\bar \cD}_{\b} {\mathfrak H}_{\a(2s)}
\non \\
&&-  \frac{2s-1}{2} \Big( \mathbb{V}^{\a(2s-2)} \cD^{\b \g} {\mathfrak H}_{\b \g \a(2s-2)}
+ \frac{1}{2}
\mathbb{V}^{\a(2s-2)} \mathbb{V}_{\a(2s-2)} \Big) \non\\
&&+2s(2s-1) \cS \mathfrak{V}^{\a(2s-2)} \mathbb{V}_{\a(2s-2)}
\non \\
&& 
+\frac{1}{8}(s-1)(2s-1)\Big(
\cD_{\b} \mathfrak{V}^{\b \a(2s-3)} \bar \cD^2 \cD^{\g} \mathfrak{V}_{\g \a(2s-3)}
+{\rm c.c.} \Big)
\bigg\}~.
\label{action2}
\eea
This action involves the real linear field strength
\bea
 \mathbb{V}_{\a(2s-2)} = \ri \cD^{\b} \bar \cD_{\b} \mathfrak{V}_{\a(2s-2)} ~,
\eea
which is invariant under \eqref{prep-gauge2}.
It varies under the transformation \eqref{4.10} as 
\bea
\d_\l {\mathbb{V}}_{\a(2s-2)} &=& \frac{\ri}{4s}
\big( \cD^{\b} \bar \cD^2 \l_{\b \a(2s-2)}+ \bar \cD^{\b} \cD^2 \bar \l_{\b \a(2s-2)} \big)~.
\label{bbV-gauge}
\eea
Setting $s=1$ in \eqref{action2} gives the type III minimal supergravity 
action in (2,0) AdS superspace, which was originally derived 
in section 10.2 of \cite{KT-M11}.\footnote{Type III supergravity is known only at the 
linearised level. In the super-Poincar\'e case, it is a 3D analogue of the 
massless superspin-3/2 multiplet proposed in \cite{BGLP}.}


\section{Discussion}

In this paper we did not carry out a systematic analysis 
(similar to that given by  Dumitrescu and Seiberg \cite{DS} 
for ordinary supercurrents in Minkowski space)
of the higher-spin supercurrent \eqref{1.1}.
The explicit form of the multiplet of currents 
was deduced from the consideration of simple dynamical systems 
in (2,0) AdS superspace.
However, the formal consistency of \eqref{1.1} follows from 
the structure of the massless higher-spin gauge theories constructed 
in section \ref{section4}.
For instance, within the framework of
the type II formulation let us couple the prepotentials 
${\mathfrak H}_{ \a (2s)} $ and ${\mathfrak L}_{ \a (2s-2)} $
to external sources
\bea
S^{(s+\hf)}_{\rm source}= \int \rd^3x \rd^2 \q  \rd^2 \bar \q \, E\, \Big\{ 
{\mathfrak H}^{ \a (2s)} J_{ \a (2s)}
-2 {\mathfrak L}^{ \a (2s-2)} {\mathbb Z}_{ \a (2s-2)}
 \Big\}~.
\eea
Requiring $S^{(s+\hf)}_{\rm source}$ 
to be invariant under the gauge transformations 
\eqref{prep-gauge} tells us that the real supermultiplet ${\mathbb Z}_{\a(2s-2)}$
is covariantly linear, 
\bea
\bar \cD^2 {\mathbb Z}_{\a(2s-2)}=0~. 
\eea
If we also require $S^{(s+\hf)}_{\rm source}$ to be invariant under the gauge transformations
\eqref{lambda-gauge}, we obtain the conservation equation
\bea
\bar \cD^\b J_{\b \a_1 \dots \a_{2s-1}} 
= \ri \bar \cD_{(\a_1}  {\mathbb Z}_{\a_2 \dots \a_{2s-1})}~.
\eea
Additionally, taking the type III formulation into account leads to the  
general conservation equation 
\bea\label{ce1}
\bar \cD^{\b} J_{\b \a(2s-1)} = \bar \cD_{(\a_1} \big( \mathbb{Y}_{\a_2 \dots \a_{2s-1})} + \ri \mathbb{Z}_{\a_2 \dots \a_{2s-1})} \big)~,
\eea
where the real trace supermultiplets $\mathbb{Y}_{\a(2s-2)} $ and $\mathbb{Z}_{\a(2s-2)} $
are covariantly linear.

An improvement transformation exists for the higher-spin  supercurrent multiplet \eqref{1.1}.
Let us introduce
\begin{subequations}\label{5.5}
\bea
\widetilde{J}_{\a(2s)}&:=& J_{\a(2s)}+ [\cD_{(\a_1}, \bar \cD_{\a_2}] \mathbb{S}_{\a_3 \dots \a_{2s})} 
+ 2 \cD_{(\a_1 \a_2} \mathbb{R}_{\a_3 \dots \a_{2s})}~, \\
\widetilde{\mathbb{Y}}_{\a(2s-2)}&:=& 
\mathbb{Y}_{\a(2s-2)}- {\ri} \cD^{\g} \bar \cD_{\g} \mathbb{R}_{\a(2s-2)} 
+ 4(s+1) \cS \mathbb{R}_{\a(2s-2)}
\non
\\
\qquad \qquad &&+ \frac{2}{s}(s-1)\cD^{\b}\,_{(\a_1}\mathbb{R}_{\a_2 \dots \a_{2s-2}) \b} ~, \\
\widetilde{\mathbb{Z}}_{\a(2s-2)}&:=& \mathbb{Z}_{\a(2s-2)} - \ri \frac{s+1}{s} 
\cD^{\g} \bar \cD_{\g} \mathbb{S}_{\a(2s-2)} - 4 (s+1)\cS \mathbb{S}_{\a(2s-2)}
\non \\
\qquad \qquad &&- \frac{2}{s}(s-1)\cD^{\b}\,_{(\a_1}\mathbb{S}_{\a_2 \dots \a_{2s-2}) \b} ~,
\eea
\end{subequations}
with $\mathbb{S}_{\a(2s-2)}$ and $\mathbb{R}_{\a(2s-2)}$ real linear superfields.
One may check that $\widetilde{J}_{\a(2s)}, \widetilde{\mathbb{Y}}_{\a(2s-2)}$ 
and $\widetilde{\mathbb{Z}}_{\a(2s-2)}$ obey the conservation equation and 
constraints described by 
\eqref{1.1}.
In the $s=1$ case, we reproduce the result given in section 10.4 of \cite{KT-M11}. 

There is one special feature of the supergravity case, $s=1$, 
for which the supercurrent conservation equation takes the form \cite{KT-M11}
\bea
\bar \cD^{\b} J_{\b \a} = \bar \cD_{\a} \big( \mathbb{Y} + \ri \mathbb{Z} \big)~,
\label{5.6}
\eea
with the real trace supermultiplets $\mathbb{Y}$ and $ \mathbb{Z} $ being covariantly 
linear. Building on the thorough analysis of  \cite{DS},
it was pointed out in \cite{KT-M11} that there exists a well-defined improvement 
transformation that results with $\mathbb{Y} =0$. For all the supersymmetric 
field theories in (2,0) AdS superspace considered in \cite{KT-M11}, 
the supercurrent is characterised by the condition $\mathbb{Y} =0$.
Actually, this condition is easy to explain.
The point is that every 3D $\cN=2$ supersymmetric field theory with U(1) $R$-symmetry
may be coupled to the (2,0) AdS supergravity, which implies $\mathbb{Y} =0$ upon 
freezing the supergravity multiplet to its maximally supersymmetric 
(2,0) AdS background.\footnote{There is another way to explain why $\mathbb Y$
may always be improved to zero. For simplicity, let us consider the case of $\cN=2$
Poincar\'e supersymmetry, with $D_\a$ and $\bar D_\a$ being the flat-superspace 
covariant derivatives. In Minkowski superspace eq. \eqref{5.6} implies 
$\pa^{\a\b} J_{\a\b} = \ri D^\a \bar D_\a {\mathbb Y}$, and therefore 
$\mathbb{Y}=\ri D^{\a} \bar D_{\a} \mathbb{R}$, for some real linear superfield 
$\mathbb R$. If we now apply the flat-superspace version of \eqref{5.5}
with $\mathbb S=0$, we will end up with ${\mathbb Y}=0$.
}
However, in the higher-spin case it no longer seems possible 
to improve  the trace supermultiplet $\mathbb{Y}_{\a(2s-2)} $ to vanish, 
as our analysis in section \ref{section3} indicates.

The massless models \eqref{action} and \eqref{action2} describe no propagating
degrees of freedom. However, in conjunction with the superconformal higher-spin
actions in conformally flat backgrounds proposed in \cite{HKO} they can be used to 
construct topologically massive higher-spin supersymmetric theories. 
Specifically, let us consider the following gauge-invariant models:
\begin{subequations}
\bea
S_{\rm massive}^{\rm (II)}&=& 
\k {S}_{\rm SCS} [ {\mathfrak H}_{\a(2s)}] 
+ m^{2s-1} S^{\rm (II)}_{(s+\hf)} [{\mathfrak H}_{\a(2s)} ,{\mathfrak L}_{\a(2s-2)} ]~, 
\label{5.7a} \\
S_{\rm massive}^{\rm (III)}&=& 
\k {S}_{\rm SCS} [ {\mathfrak H}_{\a(2s)}] 
+ m^{2s-1} S^{\rm (III)}_{(s+\hf)} [{\mathfrak H}_{\a(2s)} ,{\mathfrak V}_{\a(2s-2)} ]~,
\label{5.7b}
\eea
\end{subequations}
with $\k$ and $m$  dimensionless and massive parameters, respectively. 
Here 
\bea
{S}_{\rm{SCS}}
[ {\mathfrak H}_{\a(2s)}] 
= - \frac{(-1)^s}{2^{s+1}}
   \int 
  \rd^3x \rd^2 \q  \rd^2 \bar \q
   \, E\,
 {\mathfrak H}^{\a(2s)} 
{\mathfrak W}_{\a(2s) }( {\mathfrak H}) 
\label{2.34}
\eea
is the superconformal higher-spin action \cite{HKO}, with 
${\mathfrak W}_{\a(2s) }( {\mathfrak H}) = \bar {\mathfrak W}_{\a(2s) }( {\mathfrak H}) $ 
being the higher-spin super-Cotton tensor. It is the unique 
 descendant of ${\mathfrak H}_{\a(2s) }$ with the following properties:
 (i) ${\mathfrak W}_{\a(2s) }$ is invariant under the gauge transformations 
 \eqref{H-gauge}; (ii) ${\mathfrak W}_{\a(2s) }$ obeys the conservation equations
\bea
\bar \cD^\b {\mathfrak W}_{\b \a_1 \dots \a_{2s-1}}=0~, \qquad 
 \cD^\b {\mathfrak W}_{\b \a_1 \dots \a_{2s-1}}=0~.
 \eea

We believe that the higher-derivative actions \eqref{5.7a} and \eqref{5.7b}
describe the on-shell massive superspin-$(s+\hf)$ multiplets formulated 
in \cite{KNT-M}.\footnote{In the case of Minkowski superspace, 
this may be proved in complete analogy with the analysis given in \cite{KO}.} 
For a positive integer $n>0$, 
a massive  on-shell multiplet of superspin
$(n+1)/2$ is described by a real symmetric rank-$n$ spinor  
$T_{\a (n)}$ subject to the 
constraints \cite{KNT-M}
\begin{subequations}  \label{5.10ab}
\bea
\cD^\b T_{ \a_1 \cdots \a_{n-1} \b} 
= \bar \cD^\b T_{ \a_1 \cdots \a_{n-1} \b}&=&0~, \\ 
\Big( \frac{\ri}{2} \cD^\g\bar \cD_\g + m \Big) T_{\a_1 \cdots \a_n} &=&0~.
\eea
\end{subequations}
It may be shown that 
\bea
\Big(\frac{\ri}{2}\cD^\g\cDB_\g\Big)^2 T_{\a_1\cdots\a_n}
&=&
\Big( \cD^{a} \cD_{a} 
+(n+2) \ri \cS\cD^\g\cDB_\g 
-n(n+2)\cS^2 \Big) T_{\a_1\cdots\a_n}
~,~~~~~~~~~~~~
\eea
where the second term on the right can be rewritten as follows:
\bea
\frac{\ri}{2}\cD^\g\cDB_\g T_{\a_1\cdots\a_n}
&=&
\cD_{(\a_1}{}^{\g} T_{\a_2\cdots\a_n)\g}
+(n+2)\cS T_{\a_1\cdots\a_n} \ .
\eea
At the component level, the equations \eqref{5.10ab} may be shown to 
describe the on-shell
massive fields in AdS${}_3$ introduced in 
\cite{DKSS,BHRST}.

It is possible to construct Lagrangian models that lead directly to the equations
\eqref{5.10ab}, by generalising the flat-space bosonic
constructions of \cite{BHT,BKRTY}.
 Such a model is formulated in terms of a real symmetric rank-$n$ spinor 
 superfield ${\mathfrak H}_{\a(n) } $ 
\bea
{S}_{\rm{massive}}
[ {\mathfrak H}_{\a(n)}] 
= - \frac{\ri^n}{2^{\left \lfloor{n/2}\right \rfloor +1}} \frac{\k}{m} 
   \int
   \rd^3x \rd^2 \q  \rd^2 \bar \q
   \, E\,{\mathfrak W}^{\a(n) }( {\mathfrak H}) 
  \Big\{ m
+\frac{\ri}{2} \cD^\g \bar \cD_\g  \Big\} {\mathfrak H}_{\a(n)}~,
\label{5.13}
\eea
where ${\mathfrak W}_{\a(n) }( {\mathfrak H}) $ is the higher-spin 
super-Cotton tensor associated with  ${\mathfrak H}_{\a(n) } $ \cite{HKO}.
The action is invariant under gauge transformations 
\bea
\d_\l {\mathfrak H}_{\a(n) } =\bar  \cD_{(\a_1} \l_{\a_2 \dots \a_n) }
-(-1)^n\cD_{(\a_1} \bar \l_{\a_2 \dots \a_n) }~,
\label{2.28}
\eea
with the  gauge parameter $\l_{\a(n-1)}$
being unconstrained complex.
The gauge invariance of \eqref{5.13} follows from the properties that 
${\mathfrak W}_{\a(n) }( {\mathfrak H}) $ is (i) gauge-invariant; and 
(ii) transverse linear, 
$\bar \cD^\b {\mathfrak W}_{\b \a_1 \dots \a_{n-1}}=
 \cD^\b {\mathfrak W}_{\b \a_1 \dots \a_{n-1}}=0$.
 The action  \eqref{5.13} becomes superconformal in the $m \to\infty$ limit.

It is of interest to carry out $\cN=2 \to \cN=1$ AdS superspace reduction of 
the massless models \eqref{action} and \eqref{action2}.
Following \cite{KLT-M12}, we can introduce a real basis for the spinor covariant 
derivatives which is obtained by replacing the complex operators 
$\cD_\a$ and $\bar \cD_\a$ with $\de_\a^I$, where $I ={\bf1}, {\bf2}$,
 defined by  
 \bea
& \cD_\a=\frac{1}{\sqrt{2}}(\nabla_\a^{\bf 1}-\ri\nabla_\a^{\bf 2})~,~~~
  \bar \cD_\a=-\frac{1}{\sqrt{2}}(\nabla_\a^{\bf 1}+\ri\nabla_\a^{\bf 2})~.
 \eea
Defining $\nabla_a =\cD_a$, the new (2,0) AdS covariant derivatives  satisfy the 
algebra
\bsubeq
\bea
&\{\nabla_\a^I,\nabla_\b^J\}=
2\ri\d^{IJ}\nabla_{\a\b}
-4\ri \d^{IJ} \cS M_{\a\b}
+4\ve_{\a\b}\ve^{IJ} \cS J
~,
\label{2_0-alg-AdS-1}
\\
&{[}\nabla_{a},\nabla_\b^J{]}
=
\cS (\g_a)_\b{}^\g\nabla_{\g}^J
~,~~~~~~
{[}\nabla_{a},\nabla_b{]}
=
-4\cS^2 M_{ab}
~.
\label{2_0-alg-AdS-2}
\eea
\esubeq
 The graded commutation relations for the operators $\de_a$ and $\de_\a^{\bf1}$
have the following properties: 
 (i) they do not involve $\de_\a^{\bf2}$; and (ii)  they are identical
to those defining $\cN=1$ AdS superspace, ${\rm AdS}^{3|2}$, see \cite{KLT-M12}
for the details.
 These properties mean that ${\rm AdS}^{3|2}$ is naturally  embedded in
 (2,0) AdS  superspace  as a subspace. 
 The Grassmann variables of (2,0) AdS superspace, 
 $\q^\m_I =(\q^\m_{\bf 1}, \q^\m_{\bf 2} )$, may be chosen in such a way that 
 ${\rm AdS}^{3|2}$ corresponds to the surface defined by $\q^\m_{\bf 2} =0$.
 Every supersymmetric field theory in (2,0) AdS superspace 
 may be reduced to ${\rm AdS}^{3|2}$. Carrying out 
the  $\cN=2 \to \cN=1$ AdS superspace  reduction of 
the massless models \eqref{action} and \eqref{action2}
will give a new understanding of the difference between these models.
It will also uncover whether one of
the massless models \eqref{action} and \eqref{action2}
contain any new $\cN=1$ supersymmetric higher spin actions
compared with those derived in \cite{KT,KP}.
\\

\noindent
{\bf Acknowledgements:}\\
SMK is grateful to  Darren Grasso for comments on the manuscript, 
and to Gabriele Tartaglino-Mazzucchelli for pointing out important 
references.  
The work of JH is supported by an Australian Government Research 
Training Program (RTP) Scholarship.
The work of SMK is supported in part by the Australian 
Research Council, project No. DP160103633.


\appendix 

\section{(2,0) AdS identities} 
\label{AppendixA}

The Lorentz generators with two vector indices ($M_{ab} =-M_{ba}$),  one vector index ($M_a$)
and two spinor indices ($M_{\a\b} =M_{\b\a}$) are related to each other by the rules:
$M_a=\hf \ve_{abc}M^{bc}$ and $M_{\a\b}=(\g^a)_{\a\b}M_a$.
These generators 
act on a vector $V_c$ 
and a spinor $\J_\g$ 
as follows:
\bea
M_{ab}V_c=2\eta_{c[a}V_{b]}~, ~~~~~~
M_{\a\b}\J_{\g}
=\ve_{\g(\a}\J_{\b)}~.
\label{generators}
\eea

The covariant derivatives of (2,0) AdS superspace hold various identities, which can be easily derived from 
the covariant derivatives algebra \eqref{deriv20}.
We have made use of the following identities:
\begin{subequations} \label{cov-id}
\bea 
\left[\cD^\a, \bar \cD^2 \right]
&=& 4\ri \cD^{\a\b} \bar \cD_\b + 4\ri \cS  \bar \cD^\a - 8\ri \cS \bar \cD^\a J - 8\ri \cS \bar \cD_\b M^{\a \b} ~, \\
\left[\bar \cD^\a, \cD^2 \right]
&=& -4\ri \cD^{\a\b} \cD_\b - 4\ri \cS \cD^\a - 8\ri \cS \cD^\a J + 8\ri \cS \cD_\b M^{\a \b} ~,
\\
\big[\cD_a, \bar \cD^2\big] &=&0~,  \qquad \big[\cD_a,  \cD^2\big] =0~, \label{A.2c}
\eea
\end{subequations} 
where $\cD^2=\cD^\a\cD_\a$, and ${\bar \cD}^2={\bar \cD}_\a {\bar \cD}^\a$. 
These relations imply the identity 
\bea
\cD^\a \bar \cD^2 \cD_\a = \cDB_\a \cD^2 \cDB^\a ~,
\label{1.44}
\eea
which guarantees the reality of the actions considered 
in the main body of the paper.

In deriving eq. \eqref{3.2}, one may find the following identities useful. 
We start with the obvious relations
\begin{subequations}
\bea
\frac{\pa}{\pa \z^\a} {\cD}_{(2)} &=&2\ri { \z}^\b {\cD}_{\a \b}~, \\
\frac{\pa}{\pa \z^\a} {\cD}^k_{(2)} &=& 
\sum_{n=1}^k\,{\cD}^{n-1}_{(2)} \,\,  2\ri \, {\z}^\b {\cD}_{\a \b}\,\, {\cD}^{k-n}_{(2)} ~, \qquad k>1
~.\label{3.3}
\eea
\end{subequations}
To simplify eq.~\eqref{3.3}, we may push ${\z}^\b{\cD}_{\a \b}$, say,  to the left 
provided that we take into account its commutator with ${\cD}_{(2)}$:
\bea
[{ \z}^\b {\cD}_{\a \b}\,, {\cD}_{(2)}] = -4\ri \,{\cS}^2 \z_\a  { \z}^\b{\z}^\g {M}_{\b \g}~.
\label{3.4}
\eea
Associated with the Lorentz generators are the operators
\bea
{M}_{(2)} &:=& {\z}^\a {\z}^\b {M}_{\a \b}~,
\eea
where ${M}_{(2)}$ appears in the right-hand side of \eqref{3.4}.
This operator annihilates every superfield $U_{(m)}(\z) $ of the form 
\eqref{e1},
\bea
{M}_{(2)} U_{(m)} =0~.
\eea
From the above consideration, it follows that
\begin{subequations}
\bea
[{\z}^\b {\cD}_{\a \b}\,, {\cD}^k_{(2)}]\, U_{(m)} &=& 0 ~, \\
\Big(\frac{\pa}{\pa \z^\a} {\cD}^k_{(2)}\Big)U_{(m)} &=& 2\ri k\, {\z}^\b {\cD}_{\a \b}\, {\cD}^{k-1}_{(2)}U_{(m)}~.
\eea
\end{subequations}
We also state some other properties which we often use throughout our calculations
\begin{subequations}
\bea
{\cD}^2_{(1)} &=& 0 ~,\\
\big[ {\cD}_{(1)}\,, {\cD}_{(2)} \big] 
&=& 
\big[ \bar \cD_{(1)}\,, \cD_{(2)} \big] = 0~,\\
\big[ \cD^\a, \cD_{(2)} \big] &=& 2\ri \, {\cS}\, \z^\a {\cD}_{(1)} ~,\\
\big[\cD^\a, \cD^k_{(2)}\big] &=& 2\ri \, {\cS} \,k \,\z^\a \cD^{k-1}_{(2)} {\cD}_{(1)}~,\\
\big[\cD^\a, \z^\b \cD_{\a \b}\big] &=& 3 {\cS} \cD_{(1)}~.
\eea
\end{subequations}


\begin{footnotesize}

\end{footnotesize}

\end{document}